\documentclass[twocolumn]{cinc}
\usepackage{graphicx}
\usepackage{amsmath}
\begin{document}
\bibliographystyle{cinc}

\title{Trajectory Analysis of ECG Motif Dynamics in the Run-up to Sudden Cardiac Arrest}


\author {Nivedita Bijlani$^{1}$, Mauricio Villarroel$^{1}$\\
\ \\ 
 $^1$ The Podium Institute of Sports Medicine and Technology, University of Oxford, United Kingdom}

\maketitle

\begin{abstract}
{\looseness=-1
Early warning signatures of sudden cardiac arrest (SCA) remain poorly characterised in long-duration ECG. We quantified pre-event changes in ECG morphology using a motif-based trajectory framework. Holter ECGs from 23 patients with annotated SCA were analysed over non-overlapping 10\,s windows. Window-level motifs were extracted to quantify trajectories of instability, consistency, dispersion, heterogeneity, and personalised-baseline distance. Each trajectory was normalised to an early baseline using $z$-scores and aligned to ventricular fibrillation (VF) onset. Abnormal burden was defined as the proportion of windows with $z\geq3$ within a rolling 10-minute window. Median sustained abnormal burden onset occurred 5.8--8.4\,h before VF across morphological metrics. Motif dispersion showed the most consistent long-horizon detection, with sustained abnormal burden $\geq1$\,h before VF in 100\% of patients and $\geq2$\,h in 89\%. Motif consistency showed the strongest late-stage change, with 70\% median abnormal burden in the final 10\,min. Peak morphological deviation occurred $\sim$2\,h prior to VF. Our label-free framework transforms longitudinal ECG analysis from an event detection approach towards a continuous characterisation of evolving cardiac change, enabling a personalised early warning of SCA, well suited to long-duration wearable ECG monitoring.
\par}
\end{abstract}

\section{Introduction}
{\looseness=-1
Sudden cardiac arrest (SCA), the abrupt cessation of effective cardiac activity, remains a major global health challenge and a leading mechanism of sudden cardiac death (SCD) \cite{tsao2023heart}. SCD accounts for an estimated 15-20\% of deaths worldwide, while survival following out-of-hospital cardiac arrest remains below 10\% in many populations \cite{hayashi2015spectrum}. Identifying individuals at risk before SCA remains difficult. Nevertheless, disease substrates manifest as changes in cardiac electrophysiology and ECG morphology \cite{drezner2017international}. Tracking their evolution could provide earlier indicators of SCA risk.
\par}
{\looseness=-1
Long-duration Holter and wearable ECG provide an opportunity to observe such changes continuously. Current systems detect beats, rhythms, and events, summarising recordings through arrhythmia burden, HR/HRV, ST segment trends, and representative waveform excerpts. However, analysis remains event-centric, providing little information about how ECG morphology evolves in the run-up to an adverse event. Recent studies have explored temporal ECG dynamics using state-space models \cite{cosic2024global}, rhythm-morphology interactions \cite{gradowski2026novel}, and longitudinal risk analysis \cite{segar2026validation}. Yet a gap remains for continuous, interpretable tracking of ECG morphology, including deviation, stability, and variability over time.
\par}
We address this gap using representative ECG motifs to construct continuous longitudinal signatures of cardiac morphology, rather than discrete cardiac events. These annotation-free dynamic signatures quantify patient-specific morphological change, stability, and variability over time. We evaluate the approach in Holter ECG from patients with sustained ventricular tachyarrhythmia, using VF onset as the event endpoint associated with SCA.
\section{Methods}
\begin{figure*}[t]
    \centering
    \includegraphics[width=\textwidth]{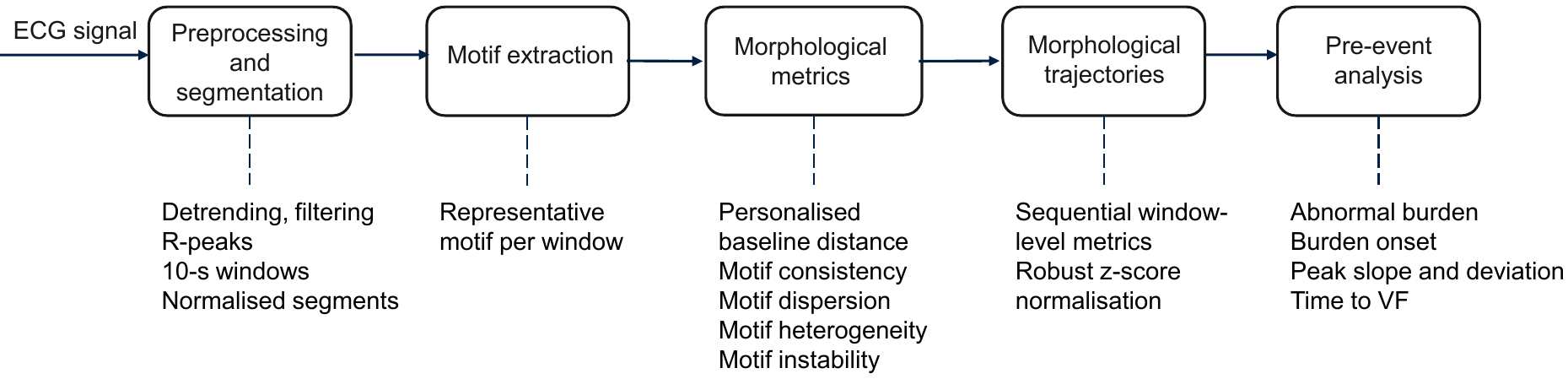}
    \caption{Overview of the motif-based framework for longitudinal ECG morphology analysis.}
    \label{fig:pipeline}
\end{figure*}

 The overall motif-based framework for longitudinal ECG morphology analysis is summarised in Fig.~\ref{fig:pipeline}.

\subsection{Dataset}
{\looseness=-1
We used the publicly available Sudden Cardiac Death Holter Database (SDDB) from PhysioNet \cite{Greenwald1986, Pollard2026PhysioNet}, which contains 23 two-channel Holter ECG recordings sampled at 250\,Hz (mean duration: $18.99 \pm 8.27$\,h). All patients experienced sustained ventricular tachyarrhythmia during monitoring, and most experienced cardiac arrest. The cohort includes 18 patients with underlying sinus rhythm, four with atrial fibrillation, and one with continuous pacing. Four sinus-rhythm patients also had intermittent pacing. Beat-level annotations are available, but are predominantly unaudited, with audited annotations limited to 12 recordings. VF onset is documented for 20 recordings and was used as the event endpoint ($t=0$); the remaining three were analysed to their endpoint. We used pre-VF ECG where available, to establish personalised baselines, construct motif trajectories, and characterise the timing of morphological changes preceding VF.
\par}

\subsection{ECG preprocessing}
We first detected R-peaks across the ECG recording. Spectral analysis showed substantial power at low frequencies (0-4\,Hz), accounting for a median 75.1\% of total signal power across recordings. To improve R-peak detection, we applied a 4th-order Butterworth band-pass filter (5-20\,Hz) and the Emrich et al. \cite{emrich2023accelerated} algorithm. We divided the ECG into consecutive 10\,s windows and band-pass filtered each at 0.5-40\,Hz for morphology analysis. Using the detected R-peaks, we extracted preceding-to-subsequent R-peak segments and temporally normalised them to 650 samples, preserving cardiac-cycle morphology while reducing heart-rate dependence. Segments were mean-centred and amplitude-normalised before motif extraction. Ectopic beats, including ventricular ectopic beats, were retained so that trajectories captured the full spectrum of morphological change.

\subsection{Motif extraction}
We extracted a representative motif, capturing recurrent cardiac-cycle morphology \cite{yeh2016matrix}, from each 10\,s window. Pairwise Dynamic Time Warping (DTW) distances were computed across R-peak-centred ECG segments:
\begin{equation}
\mathrm{DTW}(X,Y)=\min_W\sum_{(i,j)\in W}\lVert x_i-y_j\rVert ,
\end{equation}
where $W$ is the optimal warping path. DTW cost was normalised by path length and scaled to the 650-sample representation to provide a length-independent measure of morphological deviation. Within each window, we identified the segment pair with minimum DTW distance \cite{yeh2016matrix}. The segment with the lowest median distance to all other segments was selected as the representative motif. Sequential motifs captured the longitudinal evolution of ECG morphology across the recording.

\subsection{Motif-based morphological metrics}

We derived five metrics characterising morphology within and across windows:

\textit{Personalised-baseline distance (BDI)} measures change from the first 15\,min baseline as the minimum distance between each subsequent motif $\mathbf{y}_m^{(w)}$ and any baseline motif $\mathbf{y}\in\mathcal{B}$:
\begin{equation}
BDI^{(w)}=\min_{\mathbf{y}\in\mathcal{B}}
\mathrm{DTW}(\mathbf{y}_m^{(w)},\mathbf{y}).
\end{equation}

\textit{Motif consistency (MCI)} measures the median distance between the motif and all $N_w$ ECG segments within each window:
\begin{equation}
MCI^{(w)}=\mathrm{median}_{i=1,\ldots,N_w}
\mathrm{DTW}(\mathbf{x}_i^{(w)},\mathbf{y}_m^{(w)}).
\end{equation}

\textit{Motif dispersion (MDI)} measures morphological spread around the motif using the mean squared distance:
\begin{equation}
MDI^{(w)}=\frac{1}{N_w}\sum_{i=1}^{N_w}
[\mathrm{DTW}(\mathbf{x}_i^{(w)},\mathbf{y}_m^{(w)})]^2.
\end{equation}

\textit{Motif heterogeneity (MHI)} measures the motif-discord distance, where the discord $\mathbf{y}_d^{(w)}$ is the segment whose nearest-neighbour distance is maximal:
\begin{equation}
MHI^{(w)}=\mathrm{DTW}(\mathbf{y}_m^{(w)},\mathbf{y}_d^{(w)}).
\end{equation}

\textit{Motif instability (MII)} measures temporal change between consecutive window motifs:
\begin{equation}
MII^{(w)}=\mathrm{DTW}(\mathbf{y}_m^{(w-1)},\mathbf{y}_m^{(w)}).
\end{equation}

\begin{table*}[t]
\centering
\caption{Pre-VF trajectory outcomes across motif-based morphological metrics.
Timing variables are median [IQR] minutes before VF onset. Detection columns
show the percentage of recordings with burden onset at least 30, 60, or
120\,min before VF. Final burden is the percentage of abnormal windows in the
last 10\,min.}
\label{tab:trajectory_results}
\vspace{2mm}
\small
\renewcommand{\arraystretch}{1.18}
\setlength{\tabcolsep}{7pt}

\begin{tabular*}{\textwidth}{@{\extracolsep{\fill}}lccccccc@{}}
\hline
\hline
\textbf{Metric} &
\textbf{\begin{tabular}[c]{@{}c@{}}Burden onset\\(min)\end{tabular}} &
\textbf{\begin{tabular}[c]{@{}c@{}}Peak slope\\(min)\end{tabular}} &
\textbf{\begin{tabular}[c]{@{}c@{}}Peak deviation\\(min)\end{tabular}} &
\textbf{\begin{tabular}[c]{@{}c@{}}$\geq$30 min\\(\%)\end{tabular}} &
\textbf{\begin{tabular}[c]{@{}c@{}}$\geq$60 min\\(\%)\end{tabular}} &
\textbf{\begin{tabular}[c]{@{}c@{}}$\geq$120 min\\(\%)\end{tabular}} &
\textbf{\begin{tabular}[c]{@{}c@{}}Final burden\\(\%)\end{tabular}} \\
\hline

MCI & 504 [330--678] & 99 [35--162]   & 128 [69--187]  & 100 & 95  & 85 & 70 [47--92] \\
MDI & 464 [278--649] & 127 [65--189]  & 127 [73--181]  & 100 & 100 & 89 & 20 [5--36]  \\
MHI & 346 [131--562] & 204 [107--301] & 134 [62--205]  & 93  & 93  & 67 & 17 [9--25]  \\
MII & 346 [113--580] & 124 [66--182]  & 112 [60--164]  & 95  & 95  & 76 & 15 [0--29]  \\
BDI & 381 [141--622] & 113 [44--182]  & 123 [19--228]  & 94  & 88  & 75 & 12 [5--20]  \\

\hline
\hline
\end{tabular*}
\end{table*}

\begin{figure*}[t]
    \centering
    \includegraphics[width=\textwidth]{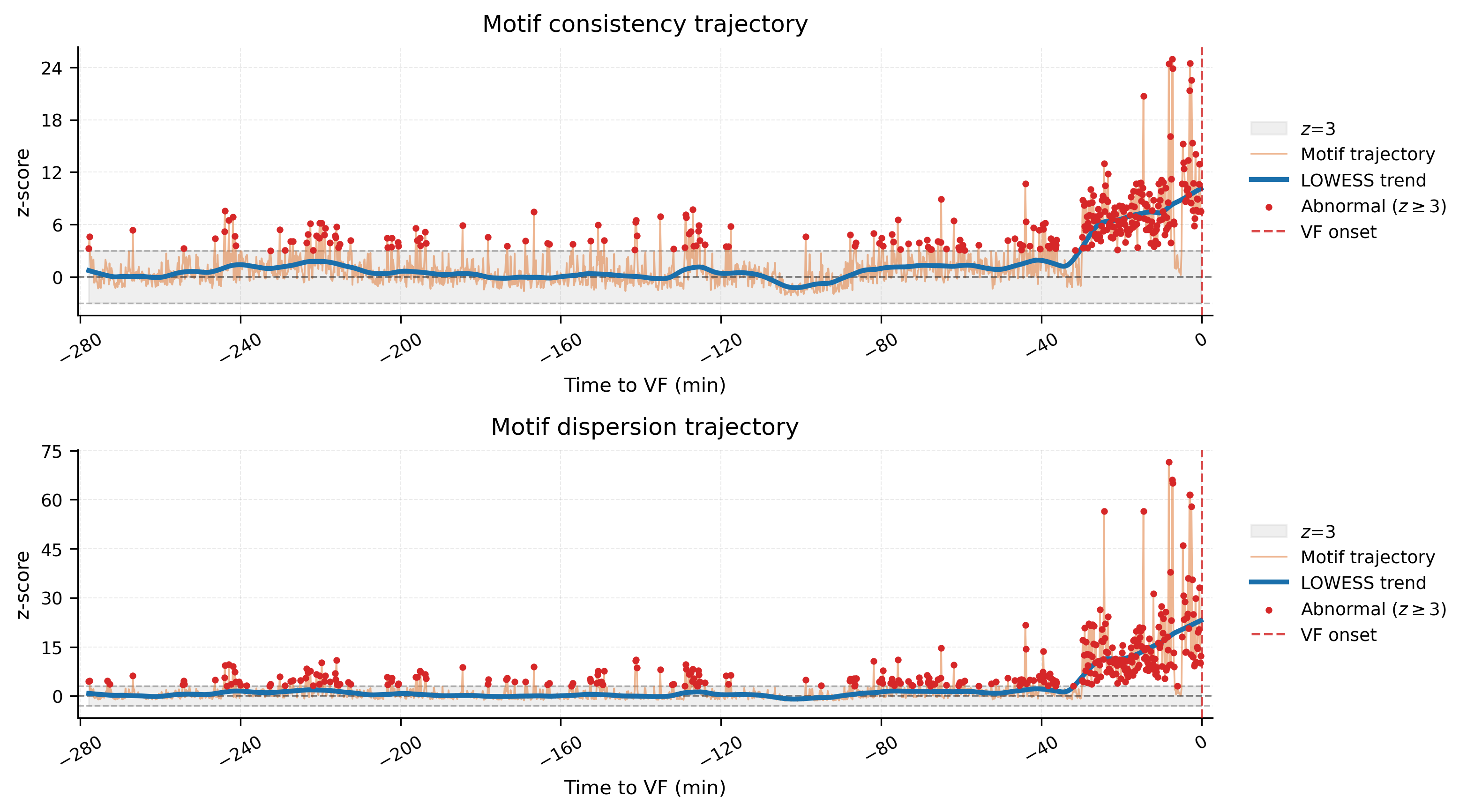}
    \vspace{2mm}
    \caption{Representative pre-VF morphological trajectories for one participant. Motif consistency (top) and motif dispersion (bottom) were normalised using robust $z$-scores relative to the 15-min baseline. Red points indicate abnormal windows ($z \geq 3$), and blue lines show LOWESS-smoothed trajectories. VF onset is defined as $t=0$.}
    \label{fig:trajectory}
\end{figure*}
\subsection{Trajectory and early-warning analysis}
{\looseness=-1
We tracked window-level motif metrics sequentially to form morphological trajectories and aligned them to VF onset ($t=0$), or to recording end where VF was not annotated. Each trajectory was normalised using $z$-scores relative to the first 15\,min baseline to account for patient-specific variability. A $z\geq3$ threshold identified outlier windows relative to the personalised baseline. We defined abnormal burden as the proportion of abnormal windows within a rolling 10\,min window, and burden onset as the first time burden reached $\geq40\%$ and persisted for at least 2\,min. Early warning was defined as burden onset occurring $\geq30$, $\geq60$, or $\geq120$\,min before VF. We also quantified the timing of maximum slope and peak deviation, and abnormal burden in the final 10\,min. 

Analysis thresholds were selected a priori to capture marked, sustained deviations across multiple warning horizons, while limiting sensitivity to transient fluctuations.
\par}
\section{Results}
Table~\ref{tab:trajectory_results} summarises pre-VF outcomes across the five motif metrics. Median burden onset occurred 5.8--8.4\,h before VF, maximum slope 1.7--3.4\,h before VF, and peak deviation 1.9--2.2\,h before VF. Sustained abnormal burden was detected $\geq30$\,min before VF in 93-100\% of event recordings, $\geq1$\,h in 88--100\%, and $\geq2$\,h in 67--89\%. MDI showed the highest detection at $\geq1$ and $\geq2$\,h (100\% and 89\%, respectively). Median abnormal burden in the final 10\,min ranged from 12--70\% across metrics.
{\looseness=-1
Fig.~\ref{fig:trajectory} illustrates the magnitude and temporal evolution of morphological change in a representative recording. MCI and MDI $z$-scores increased towards VF, with the largest increases occurring approximately 40\,min before onset.
\par}

\section{Discussion}
{\looseness=-1
Motif-based ECG trajectories captured progressive morphological changes in the hours preceding VF, with metrics reflecting complementary temporal patterns. Within-window consistency and dispersion showed the earliest sustained abnormal burden, with median onset 8.4 and 7.7\,h before VF, respectively. Motif dispersion provided the most consistent long-horizon detection (100\% at $\geq1$\,h; 89\% at $\geq2$\,h), while consistency showed the strongest late-stage change, reaching 70\% median abnormal burden in the final 10\,min. Across-window instability and personalised-baseline distance also showed sustained changes several hours before VF, with peak deviations approximately 2\,h before the event.
\par}
{\looseness=-1
These findings suggest that pre-VF morphological instability evolves across multiple temporal scales rather than through a single abrupt transition. Our framework complements event-based Holter analysis by quantifying patient-specific morphology without predefined rhythm labels or event-labelled training data. Motifs provide a compact, interpretable representation that retains a direct link to the underlying ECG, supporting lightweight longitudinal analysis of Holter and wearable recordings.

This retrospective study provides exploratory evidence to support future evaluation of SCA prediction in prospective cohorts. Sensitivity to alternative analysis settings remains to be evaluated. Future work will validate motif-based trajectories in larger cohorts, explore data-driven detection of cardiac state transitions, and develop an integrated early-warning signature.
\par}
\section{Conclusion}
{\looseness=-1

This study demonstrates how representative ECG motifs can be transformed into continuous longitudinal signatures of cardiac morphology, quantifying patient-specific change in the hours preceding VF. By modelling morphological deviation, stability, and variability over time rather than discrete events, the approach provides an interpretable, annotation-free representation of evolving cardiac state for long-duration ECG monitoring. Within athlete populations, personalised trajectories could support continuous tracking of cardiac status and help distinguish physiological adaptation from emerging pathological change. Integrated into wearable ECG, such lightweight analysis could create an actionable window for earlier clinical assessment and intervention.

\par}
\bibliography{refs}


  
  
      

\begin{correspondence}
Nivedita Bijlani\\
Mauricio Villarroel\\
The Podium Institute, Old Road Campus Research Building\\
University of Oxford, Headington, Oxford OX3 7DQ, UK\\
nivedita.bijlani@eng.ox.ac.uk\\ mauricio.villarroel@eng.ox.ac.uk
\end{correspondence}

\end{document}